\documentclass[conference,compsoc]{IEEEtran}
\ifCLASSINFOpdf
\else
\fi

\usepackage{algorithmic}
\usepackage{algorithm}
\usepackage{url}
\usepackage{cite}
\usepackage[pdftex]{graphicx}

\begin{document}
%
\title{AccuracyTrader: Accuracy-aware Approximate Processing for Low Tail Latency and High Result Accuracy in Cloud Online Services}

\author{
\IEEEauthorblockN{Rui Han}
\IEEEauthorblockA{Institute Of Computing Technology,\\
Chinese Academy of Sciences\\
Beijing, China\\
hanrui@ict.ac.cn
}
\and
\IEEEauthorblockN{Siguang Huang}
\IEEEauthorblockA{School of Software,\\
Tsinghua University\\
Beijing, China\\
huangsg15@mails.tsinghua.edu.cn}
\and
\IEEEauthorblockN{Fei Tang, Fugui Chang, Jianfeng Zhan}
\IEEEauthorblockA{Institute Of Computing Technology,\\
Chinese Academy of Sciences\\
Beijing, China\\
\{tangfei,changfugui,zhanjianfeng\}@ict.ac.cn
}
\IEEEcompsocitemizethanks{\IEEEcompsocthanksitem This paper is accepted as a Regular Paper at the 45th International Conference on Parallel Processing (ICPP-2016).\protect\\}
}

\maketitle

\begin{abstract}

Modern latency-critical online services such as search engines often process requests by consulting large input data spanning massive parallel components. Hence the tail latency of these components determines the service latency. To trade off result accuracy for tail latency reduction, existing techniques use the components responding before a specified deadline to produce approximate results. However, they may skip a large proportion of components when load gets heavier, thus incurring large accuracy losses.
This paper presents AccuracyTrader that produces approximate results with small accuracy losses while maintaining low tail latency. AccuracyTrader aggregates information of input data on each component to create a small synopsis, thus enabling all components producing initial results quickly using their synopses. AccuracyTrader also uses synopses to identify the parts of input data most related to arbitrary requests' result accuracy, thus first using these parts to improve the produced results in order to minimize accuracy losses. We evaluated AccuracyTrader using workloads in real services. The results show: (i) AccuracyTrader reduces tail latency by over 40 times with accuracy losses of less than 7\% compared to existing exact processing techniques; (ii) when using the same latency, AccuracyTrader reduces accuracy losses by over 13 times comparing to existing approximate processing techniques.

\end{abstract}


%
\IEEEpeerreviewmaketitle

\section{Introduction} \label{Section: Introduction}

Providing quick responsiveness (within 100ms) to user requests is crucial for today's online services such as e-commerce sites and web search engines, as their potential profits are proportional to service latency (request response time) \cite{farber2006google,jalaparti2013speeding}, which includes both the request queueing delay and the time of being processed.
This paper focuses on a wide class of highly parallel services, in which the processing of each request needs to consult a large input dataset by parallelizing sub-operations across hundreds or thousands of service components. Each component needs to process a subset of the input dataset to produce a result and hence the \emph{tail latency} (e.g. the 99.9th percentile latency) of these components determines the overall service latency \cite{tailatScale,li2014tales}.
Example services are: (1) \emph{services using numeric datasets}. At an e-commerce site, a user-based collaborative filtering (CF) recommender system predicts an active user's rating on an unknown item (product) by scanning millions of existing ratings from similar-minded users in a user-item rating matrix \cite{su2009survey}.
(2) \emph{Services using text datasets}. A web search engine uses an inverted index to organize millions of web pages. For each query, the search engine calculates these web pages' similarity scores to the query words (terms) and ranks the pages in descending order according to their scores.

When delivering services in a cloud platform, service providers usually have limited budgets, namely limited resources, to maintain the quality of service (QoS) requirements of their services. Hence under resource and response time constraints, a wide applied solution is to produce approximate results in request processing in order to trade off result accuracy (correctness) for service latency reduction \cite{he2012zeta,jalaparti2013speeding,agarwal2013blinkdb,chippa2013analysis}.
For example, in CF-based recommender systems and search engines, the result accuracies are the errors between predicted and actual ratings and the proportion of the actual top $k$ web pages (e.g. the top 10 pages that represent the best answers to the query terms) in the retrieved (returned) top $k$ pages, respectively \cite{chippa2013analysis}. As small accuracy losses cannot be evidently perceived and thus are tolerable by service users \cite{agarwal2013blinkdb}, efficiently and successfully applying such approximate processing mechanism requires reducing component tail latency without incurring large losses in result accuracy.

This task is difficult enough for highly distributed services deployed in a cloud environment, in which service components hosted across different nodes usually have large performance variance. This variance comes from different hardware and software reasons \cite{tailatScale} as well as frequently changing performance interference from co-located workloads such as short-running MapReduce jobs \cite{chen2012interactive,han2015pcs}. Furthermore, such performance variance is significantly amplified by request queuing delays when considering service load variations, thus incurring high component tail latency \cite{li2014tales}.
Existing techniques reduce tail latency by using results only from a part of the components responding before a specified deadline to produce approximate results \cite{he2012zeta,tailatScale,jalaparti2013speeding}. However, they do not address issues relating to reducing components' latencies themselves. This means under heavy loads, these techniques have to skip results from a large proportion of slow components to maintain low tail latency. These skipped results may cause large accuracy losses because processing the input data on all components potentially contributes to result accuracy.

In this paper, we propose AccuracyTrader, an approximate request-processing framework for low tail latency and high result accuracy in cloud online services. The basic approach taken by AccuracyTrader is to pre-create a small synopsis to aggregate the information of similar input data points on each component, and then use this synopsis to estimate the correlations between different parts of the input data and arbitrary requests' result accuracy at runtime.
AccuracyTrader thus maintains low tail latency by enabling all components producing approximate results quickly using the synopses, while still providing high result accuracy by first using the most accuracy-related input data to improve the produced results. Note that the proposed framework is not intended to replace, but rather complement the existing tail latency reduction techniques based on producing exact results \cite{tailatScale,jeon2014predictive,wu2015costlo,li2014tales,he2015reducing,han2015pcs,jalaparti2013speeding,suresh2015c3}. AccuracyTrader also differs from traditional techniques that pre-compute structures (e.g. samples or wavelets) of input data based on past query templates and use these structures to answer \emph{certain} types of requests with both accuracy and latency bounds \cite{agarwal2013blinkdb}. In contrast, AccuracyTrader needs no prior knowledge about the requests to be processed and it can support arbitrary requests in services.


We have implemented the proposed framework and modified two online services, namely a recommender system \cite{RecommenderSystem} and a web search engine \cite{luceneSearchEngine}, to adapt their request processing using AccuracyTrader to study its effectiveness. We first tested the synopsis generation and updating using real-world datasets in both services. The results show that by processing the generated synopses, the parts of input data with higher estimated correlations are indeed more related to different requests' result accuracy. We further compared AccuracyTrader against existing tail latency reduction techniques, using both the synthetic workloads in the recommender system and the realistic search engine workloads derived from the historical user queries of Sogou search engine \cite{sogouquerylogs}. The evaluation results show: (i) compared to the request reissue technique based on producing exact results \cite{jalaparti2013speeding,tailatScale,suresh2015c3}, AccuracyTrader reduces the component tail latency by more than 40 times with small accuracy losses of less than 7\%; (ii) compared to the partial execution technique based on producing approximate results \cite{he2012zeta,tailatScale,jalaparti2013speeding}, AccuracyTrader reduces the accuracy loss by more than 13 times when using the same service latency.

The remainder of this paper is organized as follows: Section \ref{Section: AccuracyTrader} introduces our approach and Section \ref{Section: Implementation} presents its implementation. Section \ref{Section: Evaluation} evaluates the proposed approach. Section \ref{Section: Related Work} discusses the related work, and finally, Sections \ref{Section: Conclusion} summarizes the work.

\section{AccuracyTrader} \label{Section: AccuracyTrader}

In this section, we first present an overview of the AccuracyTrader framework in Section \ref{Section: Overview}, following by an explanation of its modules in Sections \ref{Section: Offline Synopsis Management} and \ref{Section: Online Accuracy-driven Approximate Processing}.

\subsection{Overview} \label{Section: Overview}

\begin{figure}
\centering
  \includegraphics[scale=0.43]{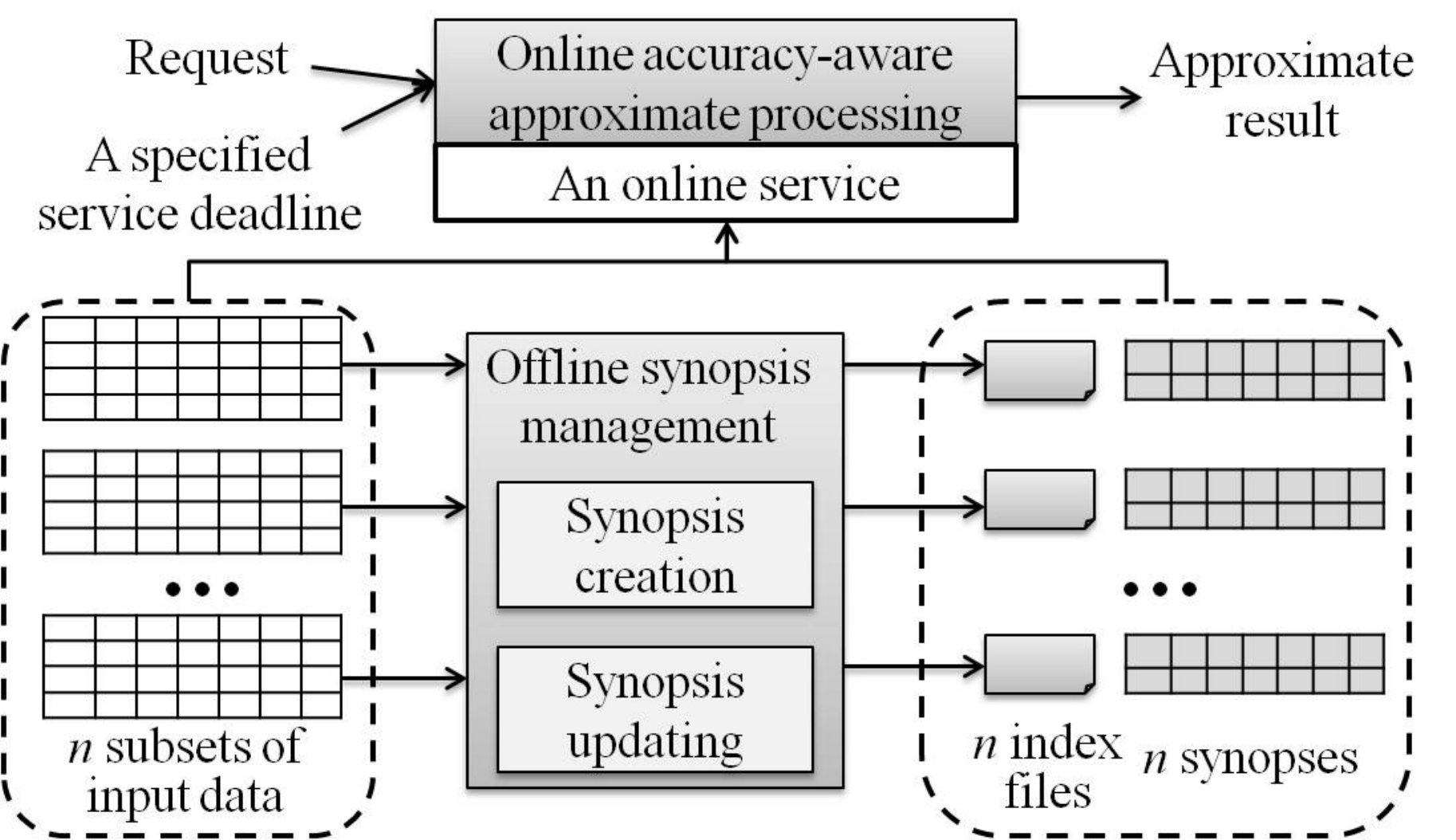}\\
  \caption{The Overview of AccuracyTrader}
  \label{Fig: framework}
\end{figure}

Suppose in an online service, the entire input data is divided into $n$ subsets for parallel processing on $n$ components.
AccuracyTrader is presented to enable the accuracy-aware approximate processing on each component using two modules, as shown in Figure \ref{Fig: framework}.


\textbf{Offline synopsis management}. This module is responsible for pre-creating and updating synopses in the offline mode. The module consists of two sub-modules. The \emph{synopsis creation} sub-module organizes each subset of input data using a proper data structure to transform the subset into an index file and a synopsis. The \textbf{synopsis} consists of multiple aggregated data points, each aggregates the information of multiple similar data points in the subset. The \textbf{index file} records the mapping relationship between each aggregated data point and the original data points aggregated by it. Note that the \emph{synopsis creation} process is only applied once, and the \emph{synopsis updating} sub-module periodically updates the created synopses in an incremental fashion to keep pace with input data changes during service running.

\textbf{Online accuracy-aware approximate processing}. For a request, this module is applied on each component to produce a result using two stages.
The first stage produces an initial approximate result using the synopsis, which is sufficiently small (e.g. 100 times smaller than the input data) such that the production process only causes a low latency even when handling heavy loads.
By processing the synopsis, this stage also estimates the correlations between different parts of the input data and the request's result accuracy.
The second stage iteratively improves the produced result within a specified service deadline. The most accuracy-related parts are first used in the improvement to minimize the request's accuracy loss.


\subsection{Offline Synopsis Management}  \label{Section: Offline Synopsis Management}

The basic idea of \textbf{synopsis creation} is to group similar data points in a subset of input data and store their aggregated information in a synopsis to preserve data similarity. In AccuracyTrader, R-tree is used in synopsis creation and updating for three reasons. First, in R-tree construction, data points close in feature attributes are allocated to the same node. Second, an R-tree is a depth-balanced tree, which means the nodes at the same depth contain similar numbers of data points and these nodes thus have the same approximation level to the subset. Third, an R-tree is an index structure that supports dynamic insertion and deletion of leaf nodes, thus enabling the incremental updating of an existing synopsis.
Based on R-tree, the synopsis creation process has three steps.

\emph{Step 1. Dimensionality reduction of the subset}. As the R-tree index model works effectively in low-dimensional spaces, this step employs the singular value decomposition (SVD) dimensionality reduction technique to transform the subset into a low-dimensional and dense dataset. SVD can transform a $u\times v$ dataset into a $u\times j$ dataset where $j$ is much smaller than $v$, while minimizing the difference (distance) between the two datasets.
AccuracyTrader uses the incremental SVD \cite{gorrell2006generalized} whose execution time is independent of the dataset size and hence the transformation process can be completed quickly (within a few seconds) even when dealing with large-scale datasets. Note that the above step works on numeric datasets. For a text dataset such as a collection of web pages, this dataset needs to be transformed into a numeric dataset, in which each data point extracts the feature attribute of its corresponding text data. For example, a web page can be transformed into a numeric data point whose attributes are all the words in the collection of web pages and the value of a attribute is the occurrence number of a word in the web page.

\emph{Step 2. Similar data points organization}. This step operates on the low-dimensional dataset and groups similar data points in it by constructing an R-tree. In the R-tree, a node including multiple data points corresponds to an \emph{aggregated data point}, and all the nodes at one depth of the tree correspond to the aggregated data points in the \emph{synopsis}.
This step outputs an index file by selecting a depth such that it contains a sufficient number of R-tree nodes to enable the fine-grained differentiation of the data points enclosed by different nodes.
The number of R-tree nodes at this depth (i.e. the number of aggregated data points in the synopsis) should also be much smaller (e.g. 100 times smaller) than the number of data points in the subset, thus guaranteeing the quick processing of the synopsis.

\emph{Step 3. Information aggregation of original data points}. According to the index file, the final step obtains each aggregated data point's corresponding \emph{original} data points (without feature reduction) and aggregates their information to generate the synopsis.
Depending on the type of dataset, there are two ways to perform such aggregation. (1) For a numeric dataset, the aggregated information can be the mean of original data points's attribute values. For example, in CF-based recommender systems, suppose an aggregated user (data point) corresponds to a set of a set $U$ of original users, in which a subset $U_i \subseteq U$ of users have rated an item $i$. The aggregated user's rating on item $i$ is users' average rating on $i$ in set $U_i$. (2) For a text dataset, the aggregated information can be the merged information of multiple data points. For example, in a search engine, suppose an aggregated web page corresponds to a set of web pages (data points), this page contains all the contents in these pages.

\begin{figure*}
\centering
  \includegraphics[scale=0.42]{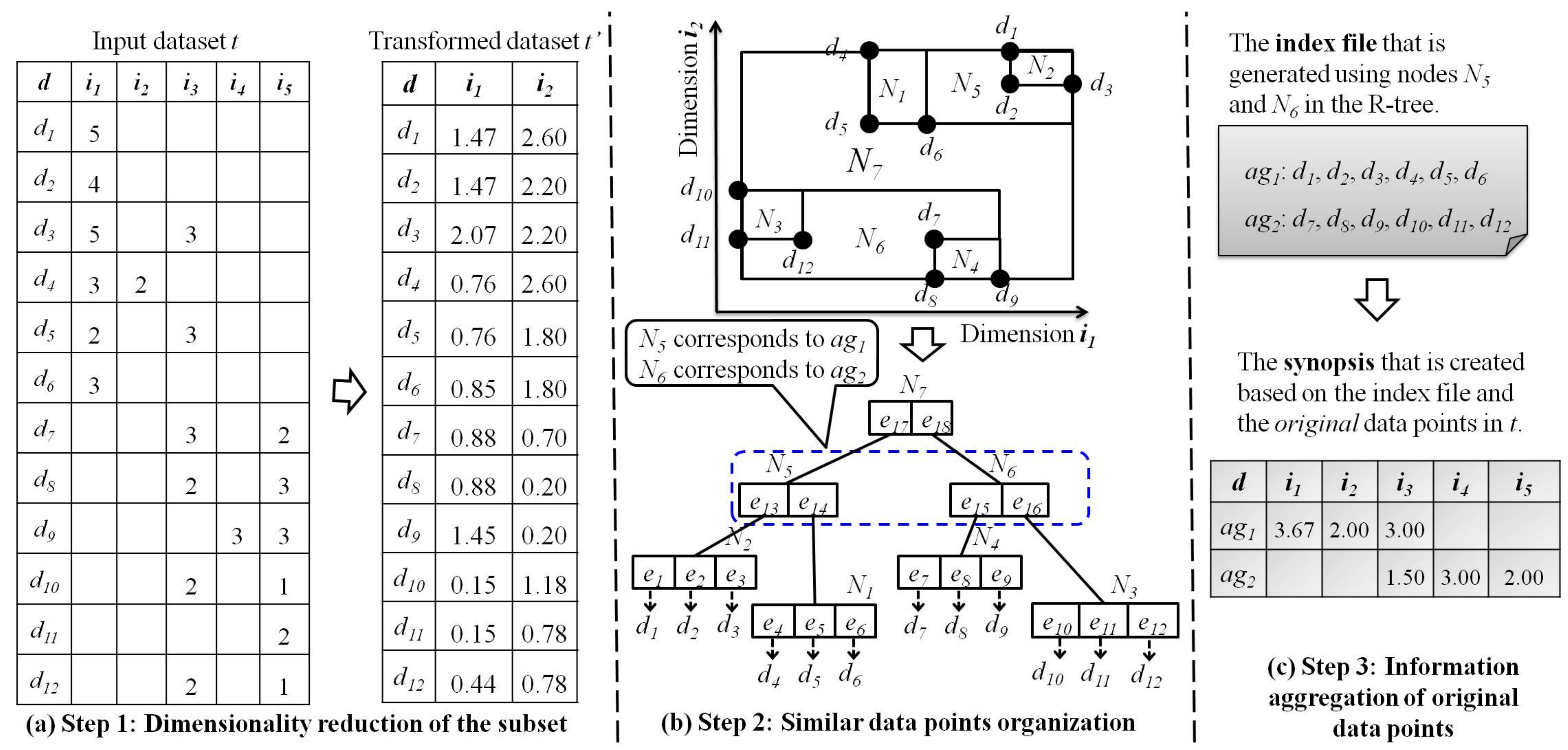}\\
  \caption{An example of offline synopsis creation}
  \label{Fig: Example}
\end{figure*}

Figure \ref{Fig: Example} shows an example process of synopsis creation. Step 1 transforms a $12\times5$ input dataset $t$ into a $12\times2$ dataset $t^\prime$. We can see that data points with similar feature attributes (e.g. points $d_1$ and $d_2$) in $t$ still have similar attributes in $t^\prime$. Step 2 organizes the 12 data points in $t^\prime$ by constructing an R-tree, in which similar data points are grouped in the same leaf node. Leaf and non-leaf nodes are then recursively grouped together following the same principle to preserve data similarity. Step 2 selects nodes $N_5$ and $N_6$ to generate an index file. Finally, step 3 creates a synopsis consisting of two aggregated data points, each aggregates information of six original data points according the index file.

Motivated by the fact that input data of online services continually changes, \textbf{synopsis updating} is designed to periodically update the existing set of synopses. To minimize the overheads in updating, this module detects changes in input data and only updates the synopsis parts influenced by the changes.
This updating strategy is built upon the dynamic insertion and deletion of leaf nodes in an R-tree and it considers two situations of input data changes. In the first situation, new input data points are added. This module thus adds new leaf nodes to incorporate these points into the R-tree.
In the second situation, the feature attributes or contents of a proportion of existing data points change. This module thus deletes the leaf nodes including these data points and inserts new leaf nodes to represent the changed points.
In both situations, synopsis updating identifies the parts of nodes influenced by the newly inserted leaf nodes and updates their corresponding aggregated data points in the synopsis.

\subsection{Online Accuracy-aware Approximate Processing} \label{Section: Online Accuracy-driven Approximate Processing}


On each component, the steps of accuracy-aware approximate processing are detailed in Algorithm \ref{AAAP}.
An initial result $ar$ is first produced using the synopsis (line 1). Our current approach uses a sufficiently small synopsis in the production to guarantee a low latency even when handling large service loads.
Applying a load-adaptive approach that dynamically selects a synopsis of a different size according to the current load is possible and it is studied in our previous work \cite{han2015sarp,han2016sarp}, but it is beyond the scope of this paper.







Estimating the \textbf{correlations} between different parts of the input data and the request's result accuracy is the key step to enable accuracy-aware request processing and this estimation is based on processing the aggregated data points in synopsis $S$ (line 1). First, processing an aggregated data point $ag_i$ gives an estimation of correlation $c_i$ between this point and the request's result accuracy. For example, in recommender systems and search engines, the correlations are the \emph{weight} between an aggregated user and an active user and the \emph{similarity score} between an aggregated web page and query terms in a request, respectively. In addition, $ag_i$ contains the aggregated information of the original data points in set $D_i$ (i.e. a part of input data) and these points have similar feature attributes. Hence, we assume a linear dependency between $c_i$ and set $D_i$'s correlation to result accuracy. That is, a higher value of $c_i$ means the accuracy improvement brought by processing the data points in $D_i$ is larger. For example, in search engines, a higher similarity score $c_i$ means the original web pages in set $D_i$ have higher similarity scores on average. Processing the web pages in $D_i$ thus has a higher probability of finding the actual top $k$ web pages and bringing a larger increase in result accuracy.






Based on the estimated correlations, the online module first ranks the aggregated data points (line 2), and then uses the ranking order of each aggregated data point to determine the ranking order of its corresponding set (line 3). Subsequently, the online module sequentially uses the ranked sets to improve result $ar$ (line 4 to 10). The improvement process iteratively executes under two conditions: (1) the elapsed time $l_{ela}$ is smaller than the specified deadline $l_{spe}$; (2) the number $i$ of the processed sets is smaller than or equal to $i_{max}$. The second condition is based on the observation that in some cases, processing the original data points in a proportion of the top ranked sets determines most of the result accuracy.
For example, in search engines, the original web pages in the top 40\% ranked sets contain over 98\% of the actual top 10 web pages for different requests. Hence, the second condition avoids the unnecessary processing of less accuracy-related data points.


\begin{algorithm}[htbp]
\caption{Accuracy-aware approximate processing on a component}
\label{AAAP}
\algsetup{
linenosize=\small,
linenodelimiter=.
}
\begin{algorithmic}[1]
\REQUIRE $ag$: an aggregated data point;\\
$c_i$: $ag_i$'s correlation to result accuracy ($1 \leq i \leq m$);\\
$D_i$: the set of original data points represented by $ag_i$; \\
$S$=\{$ag_1$,$ag_2$,...,$ag_m$\}: the synopsis with $m$ points; \\
$ar$: the approximate result; \\
$l_{spe}$: the specified deadline of service latency; \\ 
$l_{ela}$: the elapsed service time since the request submitting time; \\ 
$i_{max}$: the maximal number of sets of original data points to be processed.\\

\STATE Process $S$ to obtain the initial $ar$ and $c_1$ to $c_m$;
\STATE Rank the $m$ aggregated data points in descending order according to their correlations to result accuracy;
\STATE Obtain the ranked sets \{$D^{\prime}_1$,$D^{\prime}_2$,...,$D^{\prime}_m$\} according to the ranking orders of aggregated data points;
\STATE $i$=0; 
\STATE Obtain the current elapsed time $l_{ela}$;
\WHILE{($l_{ela} < l_{spe}$ and $i \leq i_{max}$)}
    \STATE Process original data points in $D^{\prime}_i$ to improve $ar$;
    \STATE $i$=$i+1$;
    \STATE Obtain the current elapsed time $l_{ela}$;
\ENDWHILE
\STATE Return $ar$.
\end{algorithmic}
\end{algorithm}

\section{Implementation} \label{Section: Implementation}


AccuracyTrader is implemented in Java and it is currently targeted for services running in cloud infrastructures and Linux environment. Its offline module is implemented based on open source packages of R-tree and SVD (Section \ref{Section: Offline Module}). Its online module is incorporated with two typical parallel online services: a recommender system and a search engine  (Section \ref{Section: Online Module}).


\subsection{Offline Synopsis Management Module} \label{Section: Offline Module}



AccuracyTrader currently supports R-tree based synopsis creation and updating, which operate on a service's input data and they are independent of online request processing. First of all, \emph{step 1} of \textbf{synopsis creation} is implemented based on the incremental SVD method \cite{IncrementalSVD}. This step treats the dimensionality reduction process as a gradient descent optimization problem. Suppose a $v$-dimensional dataset is transformed into a $j$-dimensional one, the time complexity of this step is O($j\times i$), where $j$ is the number of dimensions (e.g. 3) and $i$ is the number of iterations for each dimensionality.
\emph{Step 2} is implemented using the standard R-tree package \cite{RtreeCode}. Given a dataset with $k$ data points, the time complexity of constructing an R-tree is O($k \times \log k$). Finally, \emph{step 3} (i.e. information aggregation) is the most computation expensive step, whose time complexity of generating a synopsis using a $k \times v$ dataset is O($k \times v$). To accelerate the information aggregation process when dealing with large-scale datasets, we implemented a distributed version of this step running on Spark~\cite{Spark}. This implementation is based on the observation that the information aggregation process typically has a lot of iterative computations (e.g. averaging of feature attributes), which can be significantly accelerated by Spark's in-memory computing paradigm.

Once the synopsis is generated, the R-tree and the index file are stored and they can be used as the starting point of \textbf{synopsis updating}. To minimize the updating overhead, AccuracyTrader uses a low-priority strategy to perform the synopsis updating. On each service component, the synopsis updating sub-module monitors the overall resource utilization and triggers the periodic synopsis updating when the resource is underutilized, thus ensuring little interruption to the running service.
This sub-module is implemented to detect newly arrived data or changes in the original input data, dynamically updates the R-tree and the index file, and only re-generates the parts of the synopsis according to the changes in the updated index file.








\subsection{Online Accuracy-aware Approximate Processing Module} \label{Section: Online Module}

Incorporating the online module of AccuracyTrader into a service does not require any modification in the request processing algorithm, but controlling the input dataset fed to the algorithm. For each request, the synopsis is first used to produce an initial result and the ranked sets of original data points are then used to improve the result. This implementation is independent of the type of requests to be processed at runtime.


In order to test AccuracyTrader using services of different types, we incorporated its online module into a CF-based recommender system \cite{RecommenderSystem} using numeric input datasets and a Lucene web search engine \cite{luceneSearchEngine} using text input datasets. We introduce the two services as follows.

\textbf{CF-based recommender system}.
In e-commerce sites such as Amazon and eBay, the user-based CF algorithm is a predominant type of techniques applied in many recommender systems \cite{su2009survey}. In a CF-based recommendation system, a user-item rating matrix is the input dataset used for storing the user historical ratings (preference scores) for different products (items). For a request from an active user $u$, the system predicts the $u$'s rating on a target item $i$ using two steps. The first step calculates the weight (similarity) between user $u$ and any neighborhood user who has rated the same item $i$ in the matrix. One widely applied weight measure in the CF community is Pearson's correlation coefficient. The second step generates the prediction of user $u$'s rating on item $i$ by taking a weighted average of all ratings of item $i$ from user $u$'s neighborhood users.


\textbf{Lucene web search engine}.
In today's Internet, web search engines such as Google, Bing, and Baidu are the most heavily used web services and we study the open source Lucene search engine \cite{luceneSearchEngine} as an example. At the offline web page collection stage, the web crawler crawls the web pages and builds the inverted index, which includes a vocabulary containing all the words in the crawled web pages. At the online request processing stage, if a query request does not hit the query cache, the search engine scans its index file to search web pages that match the query terms in the request, and then ranks these pages according to their similarity scores to the terms. The service then returns the ranked web pages as the result, in which a small number of top ranked pages (e.g. the top 10 pages) usually stand for the answers to the query terms \cite{chippa2013analysis}.

We implemented the distributed versions of the above services based on Storm \cite{storm} (a real-time distributed processing platform), and incorporated the AccuracyTrader online module.
Using AccuracyTrader, the synopsis-based approximate processing operations only causes slightly larger time and space (memory) consumptions than the original service. This is because the synopsis is much smaller than the service input data, and the ranking of the aggregated data points in the synopsis has a polynomial computation complexity depending on the synopsis size.
 
\section{Experimental Evaluation} \label{Section: Evaluation}


In this section, we first evaluate the AccuracyTrader offline module using large datasets in real services (Section \ref{Section:Evaluation of Offline Synopsis Management}). We then compare the AccuracyTrader online module against existing tail latency reduction techniques using different experiment settings (Section \ref{Section:Evaluation of Online Accuracy-driven Approximate Processing}).

\subsection{Experimental Settings} \label{Section:Experimental Settings}
\textbf{Experiment platform}. The experiments were conducted on Xen VMs deployed across a cluster with 30 nodes. Each node has two 6-core Intel Xeon E5645 processors, 32GB of DRAM, and eight 1TB 7200RPM SATA disk drives. Each VM has 2 cores and 4GB memory. The nodes in the cluster are connected through 1Gb ethernet network cards.
The operating system of both physical machines and VMs is SUSE Linux Enterprise Server 11 SP1. The Xen, JDK versions are 4.0, 1.7.0, respectively.
The enterprise version of Storm, Alibaba JStorm \cite{JStorm}, is used. In the JStorm distribution, the versions of JStorm, Python, and Zookeeper are 0.9.6.3, 2.7.6, and 3.4.6, respectively. The version of Hadoop distribution is 1.2.1.

\textbf{Workloads}. We test two service workloads with different request arrival rates based on the implementation of AccuracyTrader on two online services in Section \ref{Section: Online Module}. We also co-locate both service workloads with Hadoop MapReduce workloads. Two types of MapReduce jobs, namely a CPU-intensive job (WordCount) and an I/O intensive job (Sort), are tested using different input data sizes ranging from 1MB to 10GB. These MapReduce workloads represent a large fraction of short-running and offline batch jobs in the cloud. The MapReduce workloads are generated using BigDataBench-MT \cite{han2015bigdatabench}, a benchmark tool to replay workloads according to real-world traces. The arrival pattern (that is, jobs' submitting time, type, and input data) of MapReduce workloads follows the Facebook production trace provided by Statistical Workload Injector for MapReduce (SWIM) \cite{chen2012interactive,swim}.





\textbf{Compared techniques}. The \emph{basic} approach without any tail latency reduction techniques and two state-of-the-art latency reduction techniques are compared: (1) \emph{request reissue} \cite{jalaparti2013speeding,tailatScale,suresh2015c3}. If some sub-operations of a request have been executed for more than a high percentile of the expected latency for this class of sub-operations, a replica of each straggling sub-operation is sent and only the quicker replica is used. The percentile is set to 95th in our evaluations.
(2) \emph{Partial execution} \cite{he2012zeta,tailatScale,jalaparti2013speeding}. For each request, this technique only uses a part of sub-operations that complete before a specified deadline to produce its approximate result and skips other sub-operations.



\textbf{Evaluation metrics}. Both performance and accuracy metrics are used to evaluate the online services. The \emph{performance} metric is the 99.9th percentile latency of parallel components for each request. This latency also determines the request's overall service latency. The \emph{accuracy} metric is the percentage of accuracy losses, which denotes the percentage of decreased accuracies in approximate results when comparing to accuracies in exact results that are produced using full computation over the entire input data.

In recommender systems, the \emph{accuracy} is measured by the root-mean-square error (RMSE) \cite{su2009survey}, which denotes the errors between the predicted and actual values of ratings. Formally, RMSE is a weighted average error that measures the prediction accuracy for all the target items in a test set $T$: $RMSE = \sqrt{\frac{\sum_{i \in T}(p(u,i)-r_{u,i})^2}{n_T}}$, where $n_T$ represents the number of items in set $T$, $p(u,i)$ is the item $i$'s predicted rating and $r_{u,i}$ is its actual rating. In search engines, the \emph{accuracy} is measured by the proportion of the actual top 10 web pages (i.e. the 10 pages with the highest similarity scores when searching all web pages) in the retrieved top 10 pages \cite{he2012zeta,chippa2013analysis}.




\subsection{Evaluation of Offline Synopsis Management} \label{Section:Evaluation of Offline Synopsis Management}



The evaluations in this section first show the overheads of the synopsis generation and updating using the input datasets of both service workloads, and then test of effectiveness of the generated synopses.



\textbf{Evaluation of overheads of synopsis creation and updating}.
The CF-based recommender system uses the MovieLens dataset \cite{movieLens} as the input data.
In the Lucene search engine, the input dataset is the inverted index created by crawling the Sogou web page collection \cite{sogouwebpages}. The input dataset in both services is divided into 108 subsets. In the recommender system, each subset has approximately 4,000 users, 1000 items, and 0.27 million ratings. In the search engine, each subset has 0.5 million web pages to be searched.

\emph{Synopsis creation}. We tested the three steps of the synopsis generation on one node. At step 1, a subset is transformed to a 3-dimensional dataset. In SVD transformation, each dimension has 100 iterations. At step 2, the 3-dimensional dataset is organized using an R-tree to generate an index file. In the recommender system, each aggregated user corresponds to an average of 133.01 original users. In the search engine, each aggregated web page  corresponds to an average of 42.55 original pages. At step 3, the information of the subset is aggregated to generate a synopsis. For the recommender system and the search engine, a synopsis was created within 30 seconds and 40 minutes, respectively. Note that in the experiments that follow, the above subsets and the generated synopses will be used.

\begin{figure}
\centering
  \includegraphics[scale=0.42]{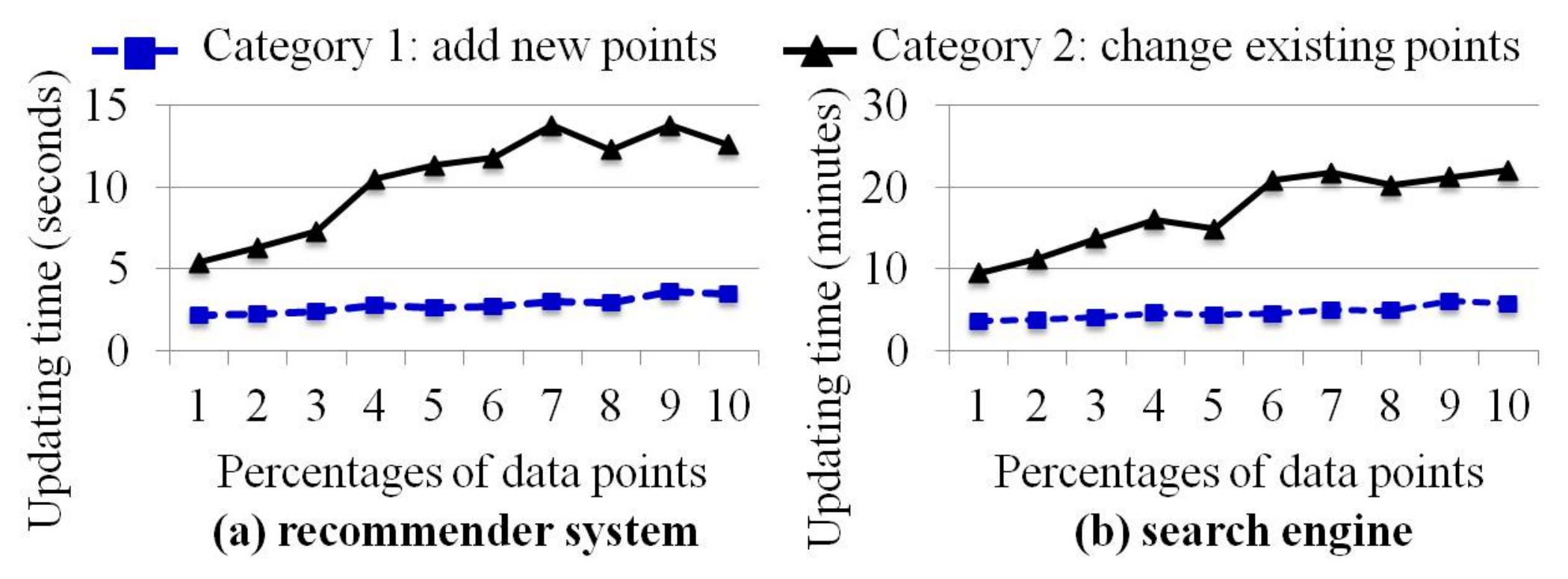}\\
  \caption{Evaluation of synopsis updating with AccuracyTrader}
  \label{Fig: synopsisUpdating}
\end{figure}

\emph{Synopsis updating}.  We designed two categories of scenarios to evaluate how the synopsis is updated under different changes in the input dataset. In the first category, each subset has a proportion of $i$\% new data points (users or web pages) being added. In the second category, each subset has a proportion of $i$\% existing data points being changed. In each scenario, 10 values of $i$ ($i$=1, 2,..., 10) were tested on one node. Each test was repeated 10 times for consistency and the average is reported in Figure \ref{Fig: synopsisUpdating}.
The evaluation results show: (i) all the updating processes were completed much faster than the synopsis creation processes; (ii) the first category of synopsis updating was completed faster than the second category.
This is because in the incremental updating of a synopsis, the first category of scenarios only needs to add new R-tree nodes. In contrast, the second category of scenarios needs to delete existing nodes and add new R-tree nodes, thus leading to longer updating time.



\textbf{Evaluation of of effectiveness of synopses}. In the AccuracyTrader framework, an aggregated data point corresponds to multiple original input data points and represents an approximation of them. This evaluation discusses whether the aggregated data points with higher estimated correlations to different requests' result accuracy really correspond to the original data points that are more related to these requests result accuracies.



\emph{Evaluation settings}. In the recommender system, the data points are users and the request are active users. An \emph{aggregated} data point's correlation to result accuracy is denoted by the weight (i.e. Pearson's correlation coefficient) between an active user and an aggregated user. An \emph{original} user is viewed as being highly related to a request's result accuracy if the weight between the active user and this user is larger than 0.8 or smaller than -0.8 (the weight ranges between -1 and 1). In the search engine, the data points are web pages and the request are queries. An \emph{aggregated} data point's correlation to result accuracy is denoted by an aggregated web page's ranking order to a query. An \emph{original} web page is viewed as being highly related to a request's result accuracy if this page belongs to the query's actual top 10 web pages.
In this evaluation, we randomly selected 1,000 active users (80\% of each user's randomly selected ratings are used in weight calculation) and 1000 queries to represent different requests.

\begin{figure}
\centering
  \includegraphics[scale=0.43]{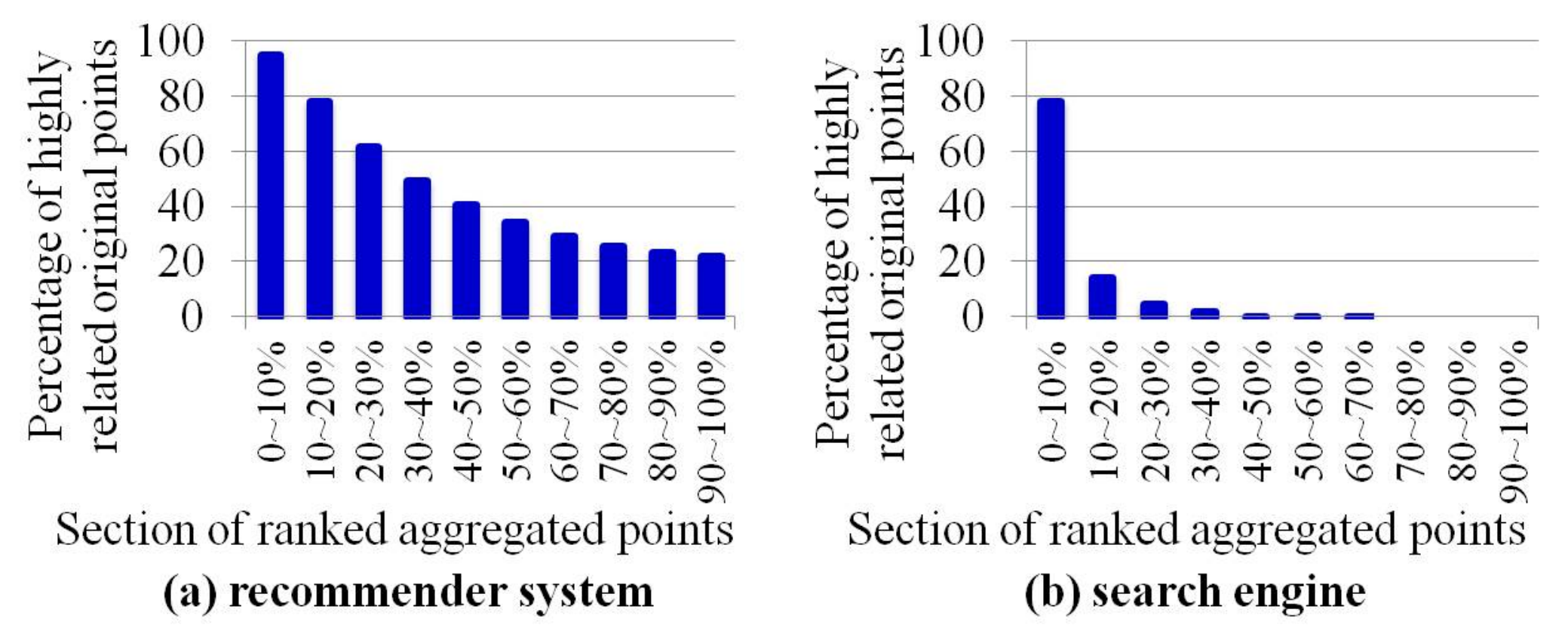}\\
  \caption{Evaluation of identifying highly related original data points with synopses}
  \label{Fig: indentifying}
\end{figure}

\emph{Evaluation results}. In Figure \ref{Fig: indentifying}, the x axis lists the \emph{ranked} aggregated data points divided into 10 sections, and the y axis shows each section's average percentage of highly related original data points when testing 1000 requests.
In the recommender system, the aggregated users are ranked and divided according to their weights to the requests. We can see in Figure \ref{Fig: indentifying}(a) that the percentage of highly related original users is 95.03\% in the first section, this percentage gradually decreases to 22.00\% in the last section.
In the search engine, the aggregated users are ranked and divided according to their ranking orders. Figure \ref{Fig: indentifying}(b) shows each section's average percentage of original web pages that are the actual top 10 web pages for the 1,000 queries. We can see that the first four sections contain 78\%, 14.17\%, 4.33\%, and 1.67\% of the actual top 10 web pages respectively, and this percentage is less than 1.17\% in the remaining six sections.

\textbf{Results}. \emph{Using the synopses, the aggregated data points with higher ranks indeed correspond to parts of input data more related to different requests' result accuracy}.



\subsection{Evaluation of Online Approximate Processing} \label{Section:Evaluation of Online Accuracy-driven Approximate Processing}

The evaluations in this section compare AccuracyTrader with existing techniques under the same deployment setting.

\emph{Deployment settings}. We deployed the service (either the recommender system or the search engine) in a cluster of 110 VMs, each hosts a service component. The 110 service components include one component for accepting and partitioning requests, 108 parallel components for processing the 108 subsets of input data, and one component for composing results to produce responses to end users. The components deployed on each node co-locate with a VM running the Hadoop MapReduce workloads to reflect the changing performance interferences.

\emph{Comparison settings}. For the \emph{basic} approach and \emph{request reissue} that produce exact results, we compare the performance between them and AccuracyTrader. We also show the accuracy losses of the approximate results produced by AccuracyTrader.
For \emph{partial execution} that produces approximate results, we set the same service latency deadline for both techniques and compare their accuracy losses.

\textbf{Comparison using the synthetic CF-based recommendation workloads}. Five request arrival rates, namely 20, 40, 60, 80, and 100 requests/second, were tested. For each test, we randomly selected 1,000 users as the active users from the Movielens dataset. For each active user, we further randomly selected 20\% of items to predict their ratings. AccuracyTrader is set to process as many original data points as possible within the specified deadline because all these points potentially contribute to result accuracy according to Figure \ref{Fig: indentifying}(a).

\emph{Evaluation results}. Table \ref{table: Comparison of the 99.9th percentile component latency (ms) using the CF-based recommender workloads} shows the tail latencies of the three techniques under different request arrival rates. When the load is light (arrival rate is 20), request reissue has the smallest latency. When load gradually increases, AccuracyTrader provides the lowest tail latencies. Table \ref{table: Comparison of the request reissue technique and AccuracyTrader using CF-based recommender workloads} lists the accuracy losses of partial execution and AccuracyTrader. We can see that in all cases, AccuracyTrader causes accuracy losses of less than 5\%, and these accuracy losses are much smaller than those of partial execution.


\begin{table}[h!]
  \centering
\caption{Comparison of the 99.9th percentile component latency (ms) using the CF-based recommender workloads}
\begin{tabular}{|c|c|c|c|c|c|}
\hline
Request arrival rate & 20 & 40 & 60 & 80 & 100 \\
\hline
Basic &76&263&48186&113496&202834\\
\hline
Request reissue &63&213&13505&27599&28981\\
\hline
AccuracyTrader &87&109&118&122&130\\
\hline
\end{tabular}
\label{table: Comparison of the 99.9th percentile component latency (ms) using the CF-based recommender workloads}
\end{table}

\begin{table}[h!]
  \centering
\caption{Comparison of percentages of accuracy losses using the CF-based recommender workloads}
\begin{tabular}{|c|c|c|c|c|c|}
\hline
Request arrival rate & 20 & 40 & 60 & 80 & 100 \\
\hline
Partial execution &0.26&4.50&23.39&81.48&99.56\\
\hline
AccuracyTrader &0.08&0.70&1.59&2.69&4.82\\
\hline
\end{tabular}
\label{table: Comparison of the request reissue technique and AccuracyTrader using CF-based recommender workloads}
\end{table}


\textbf{Comparison using the realistic search engine workloads}. In this evaluation, both the query terms and their arrival patterns (that is, queries' submitting time, arrival rates, and sequences) are derived from a 24-hour user query log collected by the Sogou search engine \cite{sogouquerylogs}. AccuracyTrader is set to process at most the original data points from the top 40\% ranked aggregated data points within the specified deadline, because they include over 98.83\% of actual top 10 web pages according to Figure \ref{Fig: indentifying}(b).


We first conduct experiments using \emph{user queries of three typical hours}. As shown in Figures \ref{Fig: Comparison of the request reissue technique and AccuracyTrader using search engine workloads of hours 9, 10 and 24}(a), (e) and (i), hour 9 (i.e. 8:00 a.m. to 9:00 a.m.), hour 10, and hour 24 represent queries with increasing, steady, and decreasing arrival rates, respectively. We tested each hour separately using 60 sessions, each session lasts 1 minute and the average is reported.
Figure \ref{Fig: Comparison of the request reissue technique and AccuracyTrader using search engine workloads of hours 9, 10 and 24} demonstrates the fluctuation of tail latencies in three techniques. We can see that the basic approach (Figures \ref{Fig: Comparison of the request reissue technique and AccuracyTrader using search engine workloads of hours 9, 10 and 24}(b), (f) and (j)) causes the highest tail latencies, which become longer and longer when loads increase because the queueing time of the slowest component continuously increases. In contrast, request reissue (Figures \ref{Fig: Comparison of the request reissue technique and AccuracyTrader using search engine workloads of hours 9, 10 and 24}(c), (g) and (k)) significantly decreases tail latencies by reducing the latencies of a small proportion of the slowest components. However, this technique still causes much longer tail latencies than those of AccuracyTrader (Figures \ref{Fig: Comparison of the request reissue technique and AccuracyTrader using search engine workloads of hours 9, 10 and 24}(d), (h) and (l)) when the service is stressed by heavy workloads.
In addition, Figure \ref{Fig: Comparison of the partial execution technique and AccuracyTrader using the search engine workloads of hours 9, 10 and 24} shows that in both approximate processing techniques, the accuracy losses fluctuate with request arrival rates because heavy loads mean less input data can be processed and thus incurring larger losses. AccuracyTrader is affected less by load variations by causing much smaller accuracy losses.

\begin{figure*}
\centering
  \includegraphics[scale=0.42]{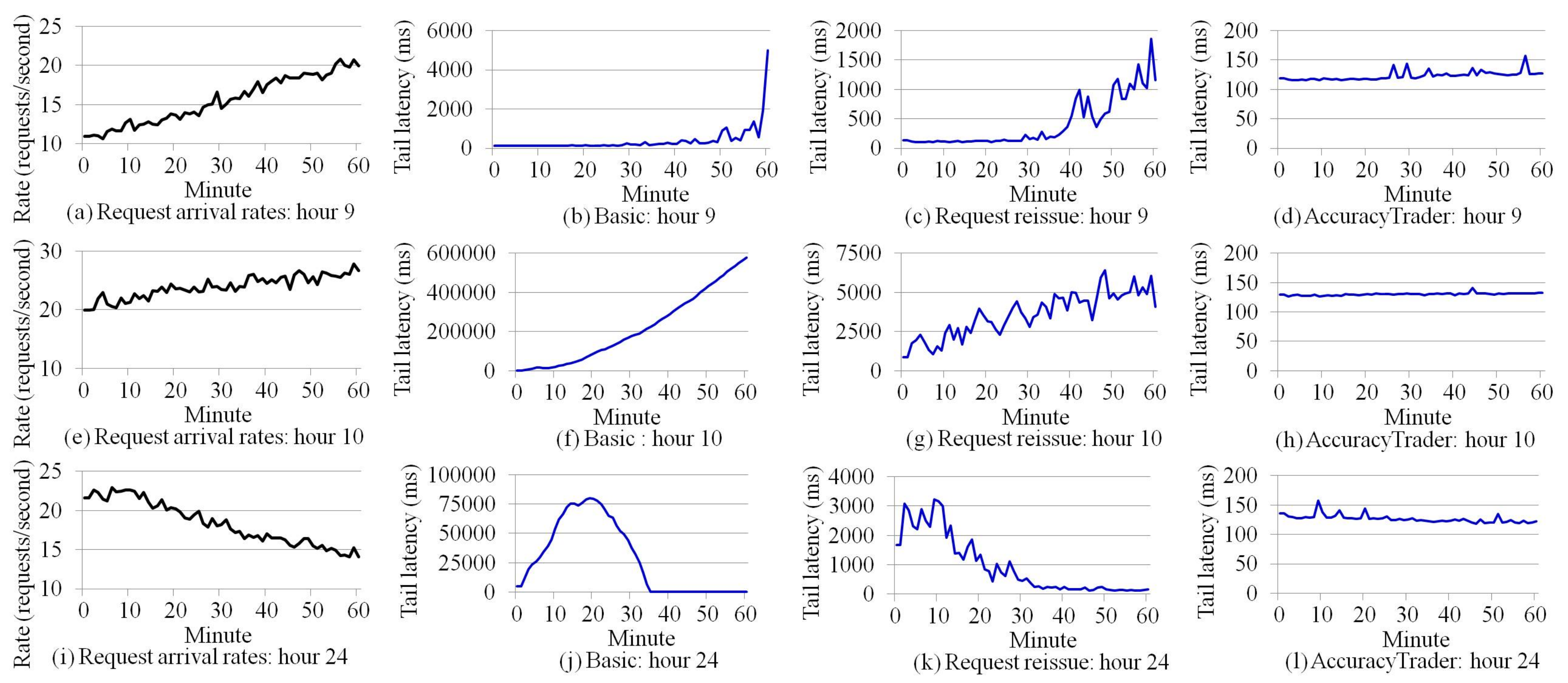}\\
  \caption{Comparison of the 99.9th percentile component latency (ms) using the search engine workloads of hours 9, 10, and 24}
  \label{Fig: Comparison of the request reissue technique and AccuracyTrader using search engine workloads of hours 9, 10 and 24}
\end{figure*}

\begin{figure}
\centering
  \includegraphics[scale=0.41]{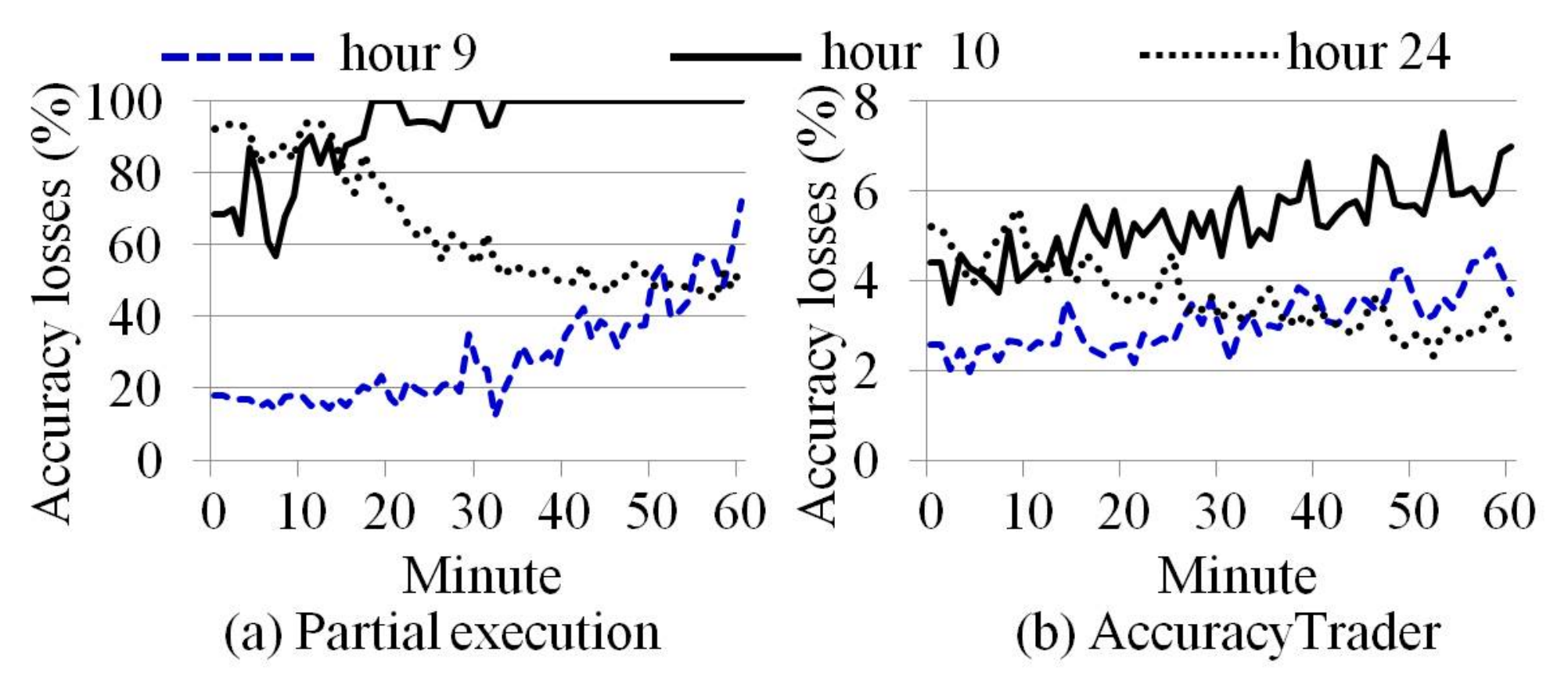}\\
  \caption{Comparison of percentages of accuracy losses using the search engine workloads of hours 9, 10, and 24}
  \label{Fig: Comparison of the partial execution technique and AccuracyTrader using the search engine workloads of hours 9, 10 and 24}
\end{figure}



\begin{figure*}
\centering
  \includegraphics[scale=0.43]{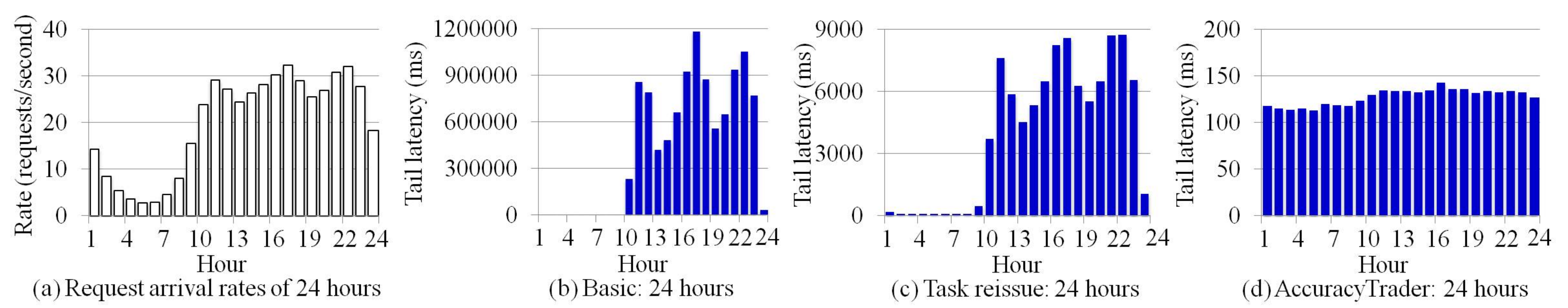}\\
  \caption{Comparison of the 99.9th percentile component latency (ms) using the search engine workloads of 24 different hours}
  \label{Fig: Comparison of the request reissue technique and AccuracyTrader using search engine workloads of 24 different hours}
\end{figure*}

\begin{figure}
\centering
  \includegraphics[scale=0.43]{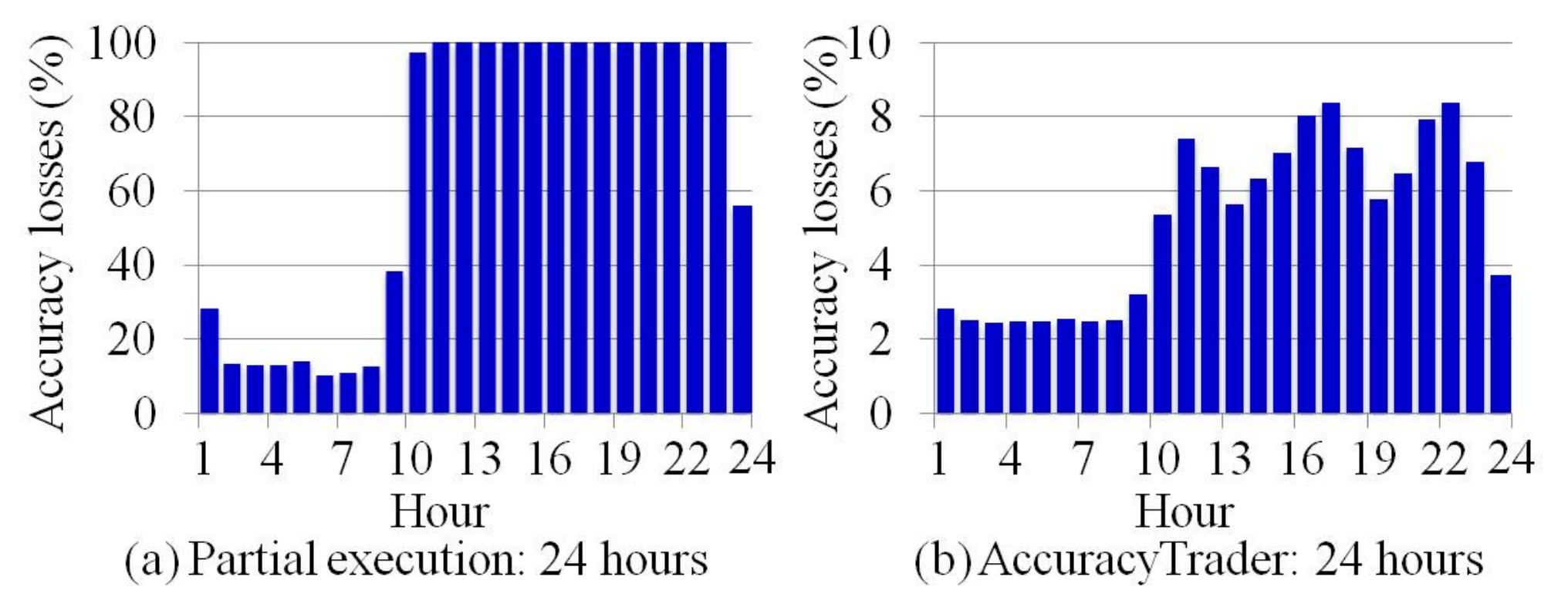}\\
  \caption{Comparison of percentages of accuracy losses using the search engine workloads of 24 different hours}
  \label{Fig: Comparison of the partial execution technique and AccuracyTrader using the search engine workloads of 24 different hours}
\end{figure}

Following the above experimental settings, we extended the comparative experiments using \emph{user queries of 24 hours a day}. Figure \ref{Fig: Comparison of the request reissue technique and AccuracyTrader using search engine workloads of 24 different hours}(a) shows the average request arrival rate of each hour. The average tail latency and accuracy loss at each hour are reported.
Figures \ref{Fig: Comparison of the request reissue technique and AccuracyTrader using search engine workloads of 24 different hours}(b), (c), and (d) show the tail latencies of three techniques. Similar to the results in previous evaluations, request reissue has the lowest tail latencies when loads are light (between hour 2 to hour 8), and AccuracyTrader has the smallest latency in other hours.  Figure \ref{Fig: Comparison of the partial execution technique and AccuracyTrader using the search engine workloads of 24 different hours} shows AccuracyTrader displays obvious superiority over partial execution by causing significantly smaller accuracy losses.

\textbf{Analysis of the compared techniques}. \emph{Request reissue} works best when load is light and parallel components have different performances. This technique thus reduces tail latency by reissuing replicas of sub-operations on straggling components to be executed on quick components. Under heavy loads, all components have high latencies because of queueing delays and this fails the reissue mechanism.
In contrast, AccuracyTrader achieves consistent low tail latencies by requiring each component completing processing within 100ms. Note that the actual latency is slightly longer than the required one. This is because AccuracyTrader has to process at least the synopsis on each component to produce a result and this processing sometimes causes longer delays than 100ms.

In \emph{partial execution}, each component still performs full computations on the entire input data. Hence under heavy loads, a majority of the components may have longer latencies than the specified deadline. This technique thus has to skip the processing results on these components and incurs large losses in result accuracy. In contrast, although AccuracyTrader only processes a small proportion of input data on each component in order to provide low latencies under heavy loads, the processed data points are the most accuracy-related ones for each request (e.g. only searching 20\% of the top-ranked web pages can find over 92\% of the actual top 10 web pages), thus only causing small accuracy losses.

\textbf{Results}. \emph{Compared to request reissue, AccuracyTrader achieves 133.38 and 42.72 times reductions in the 99.9th percentile latency with small accuracy losses of 1.97\% and 6.31\% in the evaluations of the recommender system workloads and the search engine workloads, respectively. Using the same service latency, AccuracyTrader achieves 15.12 and 13.85 times reductions in result accuracy losses compared to partial execution in the evaluations of the recommender system workloads and the search engine workloads, respectively}.
 
\section{Related Work} \label{Section: Related Work}

Reducing tail latency in highly distributed services has attracted much attentions in recent years \cite{tailatScale}. Existing techniques based on producing \emph{exact results} typically fall into three categories. The first category uses additional resources to reduce component latency variance, either by increasing the degree of parallelism \cite{jeon2014predictive} or by executing redundant requests \cite{wu2015costlo}. The second category modifies the designs of hardware and OS \cite{li2014tales} or software systems \cite{he2015reducing}. The third category mitigates the latencies of straggling components by enforcing dynamic component-node migrations \cite{han2015pcs} or reissuing requests on these components \cite{tailatScale,jalaparti2013speeding,suresh2015c3}.
Our work forms a complement to these techniques, and we do not explain them in details here. In this section, we discuss related work based on producing \emph{approximate results}.

\textbf{Approximate processing with accuracy and latency bounds}. Based on workload characteristic of past queries, some techniques pre-compute specialized structures (e.g. samples, histograms, or wavelets) of input datasets. Each structure can be used to answer a specific type of query requests with both accuracy and latency bounds \cite{agarwal2013blinkdb}. Although these techniques can provide low latency for requests with certain attributes (e.g. the high-frequency terms in search engine), they are impractical to process online services' arbitrary requests, in which the combinations of attributes are unpredictable. Hence these techniques are orthogonal to AccuracyTrader, which uses pre-computed synopses that aggregate the entire information of all attributes to support arbitrary requests.




\textbf{Partial Execution for Tail Latency Reduction}.
In large-scale online services, the partial execution technique \cite{he2012zeta,tailatScale,jalaparti2013speeding} only uses the results from a part of service components responding before a deadline to produce approximate results and skips other components.
Although providing low tail latency, all components in this technique still perform exact computations over the entire input data. Hence when loads becomes heavier, a large proportion of components cannot produce results before the deadline and the computations on the these components are skipped and wasted.
All the skipped results potentially contribute to result accuracy and this technique thus may cause large accuracy losses.
In contrast, AccuracyTrader performs computations over sufficiently small synopses to produce quick initial results on all components. Within a specified deadline, it improves the results using the parts of input data most related to result accuracy, thus resulting in high accuracy while maintaining low tail latency despite handling heavy loads.


\section{Conclusion} \label{Section: Conclusion}

In this paper, we presented AccuracyTrader, an accuracy-aware approximate processing framework for both low tail latency and high result accuracy in cloud online services.
AccuracyTrader is based on two key ideas:
(1) it aggregates information of similar input data on each component to create a small synopsis, thus enabling all components responding quickly despite handling heavy loads;
(2) it estimates the correlations between different parts of the input data and arbitrary requests' result accuracy using the synopsis, thus minimizing accuracy losses by first processing the most accuracy-related input data.
Evaluation results using both synthetic and realistic workloads demonstrate the effectiveness of AccuracyTrader at maintaining low tail latency with small accuracy losses.
 
\section{Acknowledgements}
We sincerely thank our group members, Junwei Wang, Fengming Ge, and Shulin Zhan, and the anonymous reviewers for their feedback on earlier versions of this manuscript.
This work is partly supported by National Natural Science Foundation of China (Grant Nos. 61502451), National High Technology Research and Development Program of China (Grant Nos. Y510091000), and the Key Project of of Guangdong Province, China (Grant Nos. 2015B010108006).

\bibliographystyle{plain}
\bibliography{references}

\begin{thebibliography}{10}

\bibitem{JStorm}
Alibaba jstorm.
\newblock \url{https://github.com/alibaba/jstorm}.

\bibitem{luceneSearchEngine}
Apache lucene search engine.
\newblock \url{http://lucene.apache.org/}.

\bibitem{Spark}
Apache spark.
\newblock \url{https://spark.apache.org/}.

\bibitem{storm}
Apache storm.
\newblock http://storm.apache.org/.

\bibitem{IncrementalSVD}
Incremental svd method.
\newblock \url{http://sifter.org/~simon/journal/20061211.html}.

\bibitem{RtreeCode}
Jsi (java spatial index) rtree library.
\newblock \url{http://jsi.sourceforge.net/}.

\bibitem{movieLens}
Movielens 10 million dataset.
\newblock \url{http://grouplens.org/datasets/movielens/}.

\bibitem{RecommenderSystem}
Recommender systems.
\newblock
  \url{http://www.cs.carleton.edu/cs_comps/0607/recommend/recommender/index.ht%
ml}.

\bibitem{sogouwebpages}
Sogou web pages collection.
\newblock [Online]. Available: \url{http://www.sogou.com/labs/dl/t-e.html}.

\bibitem{swim}
Statistical workload injector for mapreduce (swim).
\newblock \url{https://github.com/SWIMProjectUCB/SWIM/wiki}.

\bibitem{sogouquerylogs}
User query logs in sogou search engine.
\newblock \url{http://www.sogou.com/labs/dl/q-e.html}.

\bibitem{agarwal2013blinkdb}
Sameer Agarwal, Barzan Mozafari, Aurojit Panda, Henry Milner, Samuel Madden,
  and Ion Stoica.
\newblock Blinkdb: queries with bounded errors and bounded response times on
  very large data.
\newblock In {\em EuroSys'13}, pages 29--42. ACM, 2013.

\bibitem{chen2012interactive}
Yanpei Chen, Sara Alspaugh, and Randy Katz.
\newblock Interactive analytical processing in big data systems: A
  cross-industry study of mapreduce workloads.
\newblock {\em VLDB'12}, 5(12):1802--1813, 2012.

\bibitem{chippa2013analysis}
Vinay~K Chippa, Srimat~T Chakradhar, Kaushik Roy, and Anand Raghunathan.
\newblock Analysis and characterization of inherent application resilience for
  approximate computing.
\newblock In {\em DAC'13}, page 113. ACM, 2013.

\bibitem{tailatScale}
Jeffrey Dean and Luiz~Andr{\'e} Barroso.
\newblock The tail at scale.
\newblock {\em Communications of the ACM}, 56(2):74--80, 2013.

\bibitem{farber2006google}
Dan Farber.
\newblock Google's marissa mayer: speed wins.
\newblock {\em ZDNet Between the Lines}, 2006.

\bibitem{gorrell2006generalized}
Genevieve Gorrell.
\newblock Generalized hebbian algorithm for incremental singular value
  decomposition in natural language processing.
\newblock In {\em EACL'06}, 2006.

\bibitem{han2015sarp}
Rui Han, Junwei Wang, Fengming Ge, Jose~Luis Vazquez-Poletti, and Jianfeng
  Zhan.
\newblock Sarp: producing approximate results with small correctness losses for
  cloud interactive services.
\newblock In {\em CF'15}, page~22. ACM, 2015.

\bibitem{han2015pcs}
Rui Han, Junwei Wang, Siguang Huang, Chenrong Shao, Shulin Zhan, Jianfeng Zhan,
  and Jose~Luis Vazquez-Poletti.
\newblock Pcs: Predictive component-level scheduling for reducing tail latency
  in cloud online services.
\newblock In {\em ICPP'15}, pages 490--499. IEEE, 2015.

\bibitem{han2016sarp}
Rui Han, Jianfeng Zhan, and Jose Vazquez-Poletti Luis.
\newblock Sarp: Synopsis-based approximate request processing for low latency
  and small correctness loss in cloud online services.
\newblock {\em International Journal of Parallel Programming}, pages 1--24,
  2016.

\bibitem{han2015bigdatabench}
Rui Han, Shulin Zhan, Chenrong Shao, Junwei Wang, Jiangtao Xu, Lizy~K John,
  Lu~Gang, and Lei Wang.
\newblock Bigdatabench-mt: A benchmark tool for generating realistic mixed data
  center workloads.
\newblock {\em SoCC'15}, 2015.

\bibitem{he2015reducing}
Jun He, Duy Nguyen, Andrea~C Arpaci-Dusseau, and Remzi~H Arpaci-Dusseau.
\newblock Reducing file system tail latencies with chopper.
\newblock In {\em FAST'15}, pages 119--133. USENIX Association, 2015.

\bibitem{he2012zeta}
Yuxiong He, Sameh Elnikety, James Larus, and Chenyu Yan.
\newblock Zeta: scheduling interactive services with partial execution.
\newblock In {\em SoCC'12}, page~12. ACM, 2012.

\bibitem{jalaparti2013speeding}
Virajith Jalaparti, Peter Bodik, Srikanth Kandula, Ishai Menache, Mikhail
  Rybalkin, and Chenyu Yan.
\newblock Speeding up distributed request-response workflows.
\newblock In {\em SIGCOMM'13}, pages 219--230. ACM, 2013.

\bibitem{jeon2014predictive}
Myeongjae Jeon, Saehoon Kim, Seung-won Hwang, Yuxiong He, Sameh Elnikety,
  Alan~L Cox, and Scott Rixner.
\newblock Predictive parallelization: Taming tail latencies in web search.
\newblock In {\em SIGIR'14}, pages 253--262. ACM, 2014.

\bibitem{li2014tales}
Jialin Li, Naveen~Kr Sharma, Dan~RK Ports, and Steven~D Gribble.
\newblock Tales of the tail: Hardware, os, and application-level sources of
  tail latency.
\newblock In {\em SoCC'14}, pages 1--14. ACM, 2014.

\bibitem{su2009survey}
Xiaoyuan Su and Taghi~M Khoshgoftaar.
\newblock A survey of collaborative filtering techniques.
\newblock {\em Advances in artificial intelligence}, 2009:4, 2009.

\bibitem{suresh2015c3}
Lalith Suresh, Marco Canini, Stefan Schmid, and Anja Feldmann.
\newblock C3: Cutting tail latency in cloud data stores via adaptive replica
  selection.
\newblock In {\em NSDI'15}, 2015.

\bibitem{wu2015costlo}
Zhe Wu, Curtis Yu, and Harsha~V Madhyastha.
\newblock Costlo: Cost-effective redundancy for lower latency variance on cloud
  storage services.
\newblock In {\em NSDI'15}, 2015.

\end{thebibliography}

\end{document}